\documentstyle[aps,graphicx]{revtex}

\begin{document}
\draft
\title{Graceful exit from inflation using quantum cosmology}
\author{N. Pinto-Neto\thanks{E-mail address: {\tt nelsonpn@lafex.cbpf.br}}}
\address{Centro Brasileiro de Pesquisas F\'{\i}sicas -- Lafex\\
Rua Dr.\ Xavier Sigaud 150, Urca 22290-180 -- Rio de Janeiro, RJ -- Brazil}
\author{R. Colistete Jr.,\thanks{%
E-mail address: {\tt coliste@ccr.jussieu.fr}}}
\address{LPTL, Universit\'e Paris VI\\
Tour 22-12, 4\`eme \'etage, Bo\^{\i}te 142, 4 place Jussieu, 75252 Paris %
Cedex 05 - France.}
\date{\today}
\maketitle

\begin{abstract}
A massless scalar field without self interaction and string coupled to
gravity is quantized in the framework of quantum cosmology using the
Bohm-de Broglie interpretation. Gaussian superpositions of the quantum
solutions of the corresponding Wheeler-DeWitt equation in
minisuperspace are constructed. The Bohmian trajectories obtained
exhibit a graceful exit from the inflationary Pre-Big Bang epoch to
the decelerated expansion phase.
\end{abstract}

\pacs{PACS numbers: 98.80.Hw, 04.20.Cv, 04.60.Kz}

\section{Introduction}
\label{secIntroduction}

Superstring theory provides an exciting superinflationary model
\cite{veneziano} based on a dilaton field interaction with gravity which
is free from the common fine tuning of inflationary models because the
potential term does not play a crucial role. The superinflationary
cosmological solutions come in duality related pairs, or branches; the
first one describes a superinflationary expansion and the second branch
has a decelerated expansion. The two phases are separated by a
singularity.

The graceful exit problem of superinflation \cite{brustein} consists in
obtaining a smooth transition from the two duality solutions. It is
usually assumed that this transition can only be possible if strong
curvature effects are present (see Ref. \cite{grace} for discussion on
this point, and other possibilities).

The application of the quantum cosmology approach \cite{wdw} to the
graceful exit problem seems to be a natural choice as long as the strong
curvature regime often provides conditions for quantum cosmological
effects \cite{gaspe,lid,kiefer}. One of the main features of quantum
cosmology is the possible avoidance of classical singularities due to
quantum effects. In order to extract predictions from the wave function
of the Universe, the Bohm-de Broglie ontological interpretation of
quantum mechanics \cite{bohm,holland} has been proposed
\cite{santini,bola}, since it avoids many conceptual difficulties that
follow from the application of the standard Copenhagen interpretation to
a unique system that contains everything. In opposition to the latter
one, the ontological interpretation does not need a classical domain
outside the quantized system to generate the physical facts out of
potentialities (the facts are there {\it ab initio}), and hence it can
be applied to the Universe as a whole\footnote{Other alternative
interpretations can be used in quantum cosmology like the Many Worlds
interpretation of quantum mechanics \cite{everett}}. This interpretation
admits the concept of trajectories, called Bohmian trajectories, which
in the case of cosmology are entire histories of the Universe. These
histories satisfy a modified Hamilton-Jacobi equation, different from
the classical one due to the presence of a quantum potential term.
Hence, the Bohmian trajectories become different from the classical ones
when quantum effects become important. Recent investigations
\cite{martomoniz1,martomoniz2} have used this approach to investigate
issues of string cosmology, among them, the superinflationary behaviour
and the possibility of solving the graceful exit problem.

In the present paper we will be restricted to the simple case of no
dilaton field potential term and Friedmann-Lema\^itre-Robertson-Walker
(FLRW) geometries with flat spatial sections. The Bohmian trajectories
obtained from solutions of the Wheeler-DeWitt (WDW) equation in Jordan's
frame can be calculated through suitable conformal transformations
\cite{fabris} from the Bohmian trajectories in Einstein's frame, which
were obtained in Ref. \cite{Fla3a1} from Gaussian superpositions of
negative and positive mode solutions of the related WDW equation. The
resulting trajectories present a smooth transition from the classical
inflationary Pre-Big Bang epoch to the usual classical Pos-Big-Bang
decelerated phase. A graceful exit is created through quantum effects
which are relevant only at the transition.

In Section \ref{secClassical} we show the duality classical solutions
for superinflation in Jordan's frame. Section \ref{secQuantum} uses the
Bohmian trajectories obtained in Ref. \cite{Fla3a1} in Einstein's frame
to obtain the related trajectories in Jordan's frame. The graceful exit
transition from the superinflation Pre-Big Bang epoch to the usual
decelerated expansion phase is exhibited. The conclusions are presented
in Section \ref{secConclusion}.

\section{The classical model}
\label{secClassical}

The so called superinflation is a superstring cosmology duality-pair
solution in the low energy effective field theory context. The relevant
terms can be shown in the following Lagrangian in the Jordan frame (with
non-minimal coupling between gravity and the free scalar field):
\begin{equation}
{\it L}^{(J)}=\sqrt{-g}e^{-\phi }\biggr(R-\omega \phi _{;\rho }\phi ^{;\rho }%
\biggl) .  \label{LagJordan}
\end{equation}
This Lagrangian appears in effective string theory without the Kalb-Ramond
field when $\omega =-1$, the usual dilaton field potential $V(\phi )$ not
being considered in this paper. Through a conformal transformation
$g_{\mu \nu }=e^{\phi }\bar{g}_{\mu \nu }$, we can obtain the Lagrangian in
the Einstein frame (with minimal coupling)
\begin{equation}
{\it L}^{(E)}=\sqrt{-\bar{g}}\biggr(\bar{R}-C_{\omega }\bar{\phi} _{;\rho }
\bar{\phi} ^{;\rho }%
\biggl) ,  \label{LagEinstein}
\end{equation}
where $C_{\omega }\equiv (\omega +3/2)$. To compare with the classical
trajectories obtained in Ref. \cite{brustein}, just the $\omega =-1$ case
will be studied here, so $C_{\omega }=1/2$.

The Robertson-Walker metric with vanishing spatial curvature
\begin{equation}
\label{nense}
ds^{2}=g_{\mu \nu }{\rm d}x^{\mu }{\rm d}x^{\nu }=-N^{2}{\rm d}t^{2}+{a(t)}%
^{2}{\rm d}x_{i}{\rm d}x^{i} ,
\end{equation}
will be used in the Jordan frame. The consequence of applying the
conformal transformation on the metric (\ref{nense}) must be taken into
account when we compare the results in the Jordan and Einstein frames
(see Ref. \cite{fabris}). Then, the lapse function $N$ and the scale
factor $a$ are modified,
\begin{equation}
\label{flu}
ds^{2}=e^{\phi }\bar{g}_{\mu \nu }{\rm d}x^{\mu }{\rm d}x^{\nu }=-e^{\phi
}\bar{N}^{2}{\rm d}t^{2}+e^{\phi }{\bar{a}(t)}^{2}{\rm d}x_{i}{\rm d}x^{i} ,
\end{equation}
yielding $N=e^{\phi /2}\bar{N}$ and $a=e^{\phi /2}\bar{a}$.

Working in the gauge $N=1$ and restricting the scalar field to be time
dependent, $\phi =\phi (t)$, the solutions of the equations of motion
of the Lagrangian (\ref{LagJordan}) in the Jordan frame are given in
Ref.  \cite{brustein}, and can be organized in two branches of solutions
for $\{H\equiv \dot{a} /a,\phi \}$ as functions of time.

The $(+)$ branch, valid when $t<t_{0}$, reads
\begin{eqnarray}
H^{(+)} &=&\pm \frac{1}{\sqrt{3}}\frac{1}{t-t_{0}},  \label{Hplus} \\
\phi ^{(+)} &=&\phi _{0}+(\pm \sqrt{3}-1)\ln (t_{0}-t),  \label{phiplus}
\end{eqnarray}
and describes accelerated expansion in the case $(H>0,\dot{H}>0)$, i.e.,
inflationary evolution.

For $t>t_{0}$, the $(-)$ branch solution seems to be almost the same,
\begin{eqnarray}
H^{(-)} &=&\pm \frac{1}{\sqrt{3}}\frac{1}{t-t_{0}},  \label{Hminus} \\
\phi ^{(-)} &=&\phi _{0}+(\pm \sqrt{3}-1)\ln (t-t_{0}),  \label{phiminus}
\end{eqnarray}
but now describes decelerated expansion in the case $(H>0,\dot{H}<0)$, which
can be connected smoothly to a Friedmann-Robertson-Walker (FRW) decelerated
expansion.

The transition from an inflationary expanding universe to a FRW expanding
one, or graceful exit from inflation, will be obtained here by means of
quantum Bohmian trajectories. These trajectories are better described using
the coordinate $\alpha =\ln a$, so that the $(+)$ branch (\ref{Hplus})--(\ref
{phiplus}) in the $(H>0)$ case, gives
\begin{eqnarray}
\alpha ^{(+)} &=&-\frac{1}{\sqrt{3}}\ln (t_{0}-t),  \label{alphaplus} \\
\phi ^{(+)} &=&\phi _{0}+(3+\sqrt{3})\alpha ^{(+)},  \label{phiplus2}
\end{eqnarray}
while the $(-)$ branch (\ref{Hminus})--(\ref{phiminus}), also in the $(H>0)$
case, yields
\begin{eqnarray}
\alpha ^{(-)} &=&\frac{1}{\sqrt{3}}\ln (t-t_{0}),  \label{alphaminus} \\
\phi ^{(-)} &=&\phi _{0}+(3-\sqrt{3})\alpha ^{(-)}.  \label{phiminus2}
\end{eqnarray}

Hence, the classical trajectories in the plane $(\phi ,\alpha )$ are straight
lines with different slopes depending on the type of branch, and both have
singularities for $t=t_{0}$.

\section{Quantum cosmology with the Bohm--de Broglie interpretation}
\label{secQuantum}

In Ref. \cite{Fla3a1}, the Dirac quantization approach was applied to the
theory in the Einstein frame (\ref{LagEinstein}), giving the Wheeler-DeWitt
equation in minisuperspace whose quantum solutions
were obtained. Gaussian superpositions of these solutions given by
\begin{equation}
\label{onda}
\Psi (\bar{\alpha},\bar{\phi}) =
\int A\,F(k)\biggl[\exp (\bar{\phi} + \bar{\alpha}) +
\exp (\bar{\phi} - \bar{\alpha})\biggr] {\rm d}k ,
\end{equation}
with
\begin{equation}
F(k)\equiv \exp \biggl[-\frac{(k-d)^2}{\sigma ^2}\biggr] ,
\end{equation}
and $A,d,\sigma$ arbitrary constants, were interpreted using
the Bohm-de Broglie ontological interpretation of quantum mechanics \cite
{bohm,holland}, and the resulting system of planar equations for the case of
vanishing spatial curvature was :
\begin{eqnarray}
\frac{{\rm d}\bar{\alpha}}{{\rm d}t} &=&\frac{\bar{N}\biggl[\bar{\phi}%
\sigma ^{2}\sin (2d\bar{\alpha})+2d\sinh (\sigma ^{2}\bar{\alpha}\bar{%
\phi})\biggr]}{\exp (3\bar{\alpha})\biggl\{2[\cos (2d\bar{\alpha})+\cosh
(\sigma ^{2}\bar{\alpha}\bar{\phi})]\biggr\}} ,
\label{dalphatildedtEinstein} \\
\frac{{\rm d}\bar{\phi}}{{\rm d}t} &=&\frac{\bar{N}\biggl[-\bar{\alpha}%
\sigma ^{2}\sin (2d\bar{\alpha})+2d\cos (2d\bar{\alpha})+2d\cosh (\sigma
^{2}\bar{\alpha}\bar{\phi})\biggr]}{\exp (3\bar{\alpha})\biggl\{2[\cos
(2d\bar{\alpha})+\cosh (\sigma ^{2}\bar{\alpha}\bar{\phi})]\biggr\}} ,
\label{dphitildedtEinstein}
\end{eqnarray}
where $\bar{\alpha}=\ln \bar{a}$, and we have relabeled $\bar{\phi}%
\rightarrow \sqrt{C_w/6} \bar\phi$.

The solutions of the above equations yield the Bohmian trajectories.
Among others, there are bouncing regular solutions which contract classically
from infinity until a minimum size, where quantum effects become
important acting as repulsive forces avoiding the singularity,
expanding afterwards to an infinite size, approaching the classical
expansion as long as the scale factor increases.  For details, see Ref.
\cite{Fla3a1}\footnote{Note that the variables $(\bar{\alpha},\bar{\phi},
\bar{N})$ are the variables $(\alpha ,\phi ,N)$ of Ref. \cite{Fla3a1},
renamed to avoid misunderstanding.}. With the lapse function $\bar{N}$ we can
obtain the dependence on $t$ for $\bar{\alpha} (t)$ and $\bar{\phi} (t)$.

To obtain the Bohmian trajectories in Jordan's frame we only have to make
the following substitutions in Eqs.
(\ref{dalphatildedtEinstein})--(\ref{dphitildedtEinstein}): $\bar{\alpha}=\alpha
-\frac{\phi }{2}$ and $\bar{\phi} =\frac{\phi }{2\sqrt{3}}$ (see Eq.
(\ref{flu})). Finally, as we use $N=1$, we have $\bar{N}=e^{-\phi /2}$.
The planar system in the variables $(\alpha ,\phi )$ then reads
\begin{eqnarray}
\frac{{\rm d}\alpha }{{\rm d}t} &=&\frac{\sqrt{3}\biggl\{6d\cos [d(2\alpha
-\phi )]+6d\cosh \biggl[\frac{\sigma ^{2}(2\alpha -\phi )\phi }{4\sqrt{3}}%
\biggr]-\sigma ^{2}(3\alpha -2\phi )\sin [d(2\alpha -\phi )]\biggr\}+6d\sinh %
\biggl[\frac{\sigma ^{2}(2\alpha -\phi )\phi }{4\sqrt{3}}\biggr]}{6\exp
(3\alpha -\phi )\biggl\{\cos [d(2\alpha -\phi )]+\cosh \biggl[\frac{\sigma
^{2}(2\alpha -\phi )\phi }{4\sqrt{3}}\biggr]\biggr\}} ,
\label{dalphadtJordan} \\
\frac{{\rm d}\phi }{{\rm d}t} &=&\frac{\sqrt{3}\biggl\{4d\cos [d(2\alpha
-\phi )]+4d\cosh \biggl[\frac{\sigma ^{2}(2\alpha -\phi )\phi }{4\sqrt{3}}%
\biggr]-\sigma ^{2}(2\alpha -\phi )\sin [d(2\alpha -\phi )]\biggr\}}{2\exp
(3\alpha -\phi )\biggl\{\cos [d(2\alpha -\phi )]+\cosh \biggl[\frac{\sigma
^{2}(2\alpha -\phi )\phi }{4\sqrt{3}}\biggr]\biggr\}} .
\label{dphidtJordan}
\end{eqnarray}

\begin{figure}
\begin{center}
\includegraphics[scale=0.9]{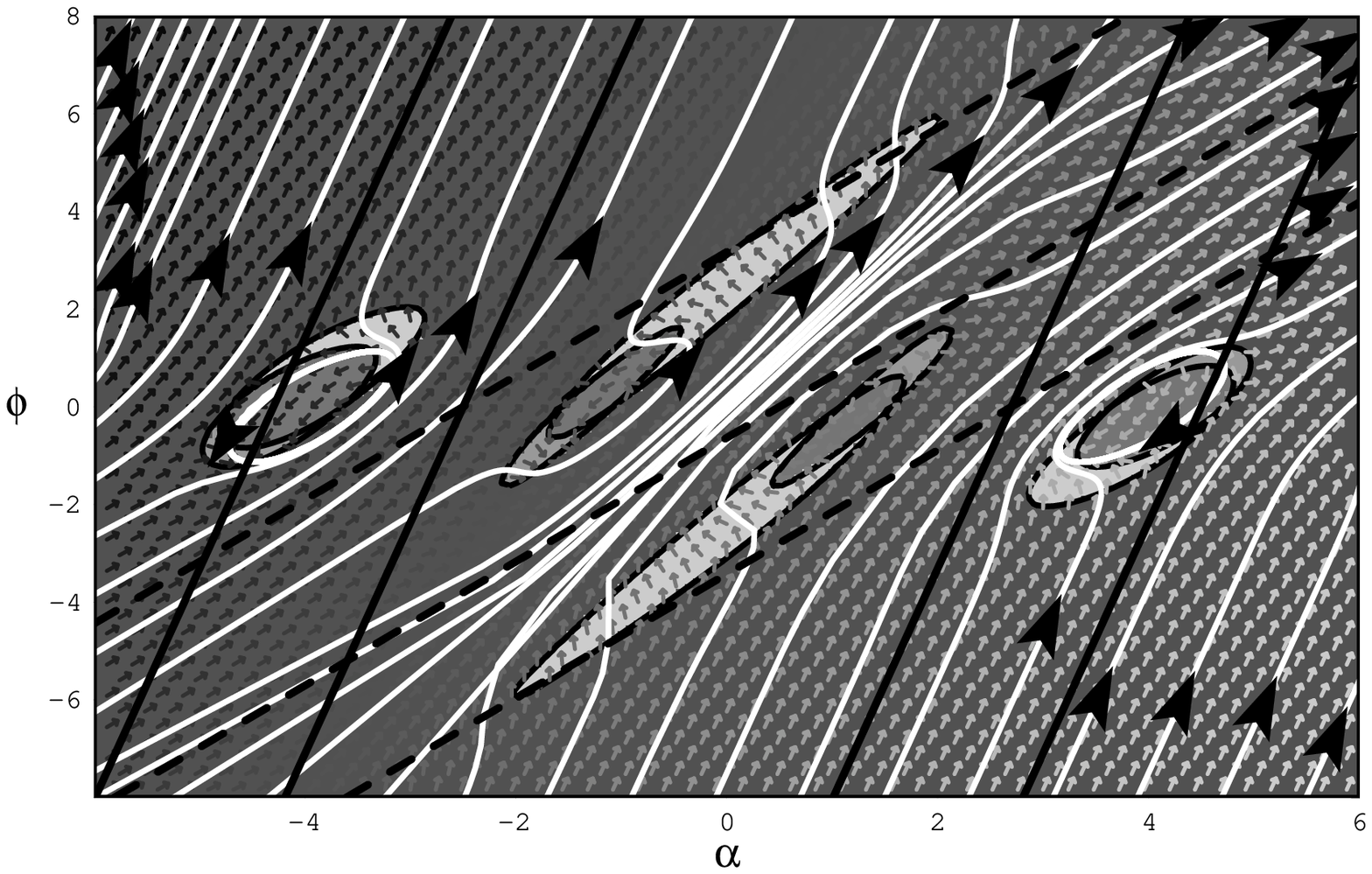}
\caption{
Field plot of the system of planar equations (\ref{dalphadtJordan})--(\ref%
{dphidtJordan}) for $\sigma =d=1$, which uses the Bohm-de Broglie
interpretation with the wave function of the Universe. Each arrow of the
vector field is shaded according to its true length, black representing
short vectors and white, long ones. The four shades of gray show the regions
where the vector field is pointing to northeast, northwest, southeast or
southwest. The black curves are the nullcline curves that separate these
regions. The trajectories are the white curves with direction arrows. The
black straight lines are parallel to the classical solutions, the solid lines
have accelerated expansion and the dashed lines have decelerated expansion.
}
\label{figphialpha}
\end{center}
\end{figure}

A field plot of this planar system is shown in Figure
\ref{figphialpha}, for $\sigma =d=1$. Depending on the initial
conditions, there are three types of behaviour for the Bohmian
trajectories. There are small oscillating universes without
singularities near the center points placed in the $\phi =0$ axis.
When $\left| (2\alpha -\phi )\phi \right| \gg 0$, Eqs.
(\ref{dalphadtJordan})--(\ref{dphidtJordan}) approach the classical
equations of motion yielding the straight lines which appear in Figure
\ref{figphialpha}, representing the classical solutions
(\ref{phiplus2})--(\ref{phiminus2}). Note, however, that in the region
of small values of $\left| \phi \right| $ there are smooth transitions
between these two straight lines which are not present in the classical
solutions. One possible type of transition is exactly the graceful
exit, from the $(+)$ branch with inflationary expansion to the $(-)$
branch solution with decelerated expansion. Another possible type of
transition is the opposite of the graceful exit, from the $(-)$
branch with decelerated expansion to the $(+)$ branch solution with
inflationary expansion.

\begin{figure}
\begin{center}
\includegraphics[scale=0.75]{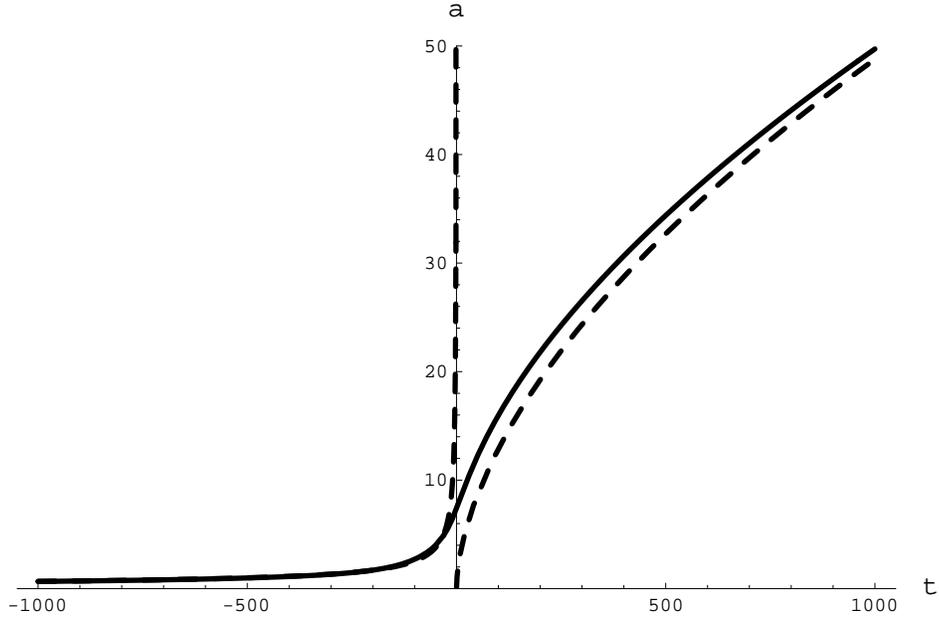}
\caption{
Plot of the scale factor $a(t)$ as function of $t$, it is one Bohmian
trajectory with initial conditions $\alpha =2$, $\phi =0$ and $t_{0}=0$
using the system of planar equations (\ref{dalphadtJordan})--(\ref{%
dphidtJordan}) for $\sigma =d=1$. The dashed lines describe the classical
solutions with accelerated and decelerated expansion which are asymptotically
obtained when $\mid t-t_{0}\mid >>0$, and there is a smooth transition near
$t=t_{0}$ avoiding the singularity.
}
\label{figa}

\end{center}
\end{figure}

\begin{figure}
\begin{center}
\includegraphics[scale=0.75]{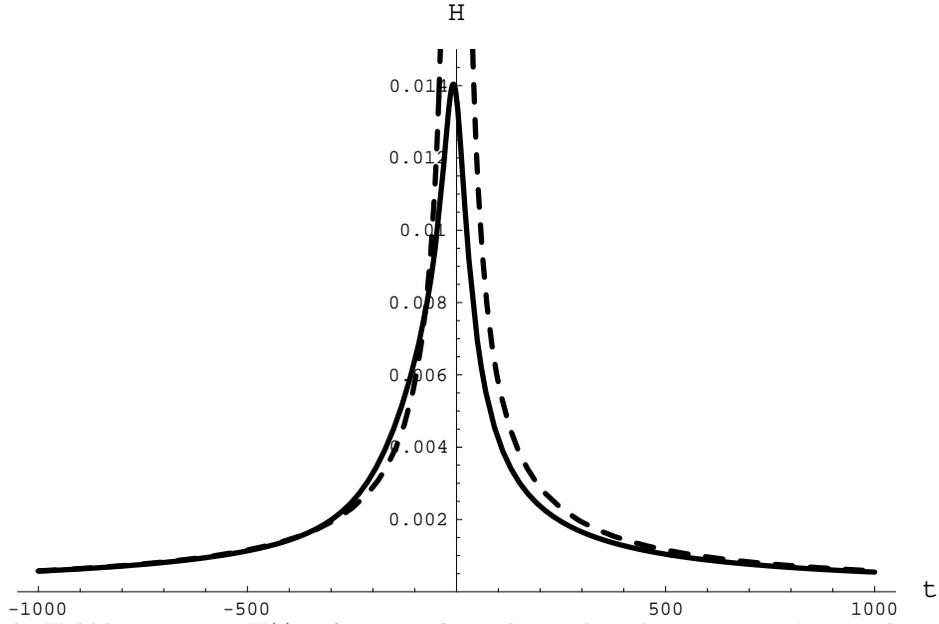}
\caption{
Plot of the Hubble parameter $H(t)$ as function of $t$, with initial conditions
$\alpha =2$, $\phi =0$ and $t_{0}=0$ using the system of planar equations
(\ref{dalphadtJordan})--(\ref{dphidtJordan}) for $\sigma =d=1$. The classical
solutions of the $(+)$ and $(-)$ branches, shown in dashed lines, are
asymptotically obtained when $\mid t-t_{0}\mid >>0$, and the smooth transition
near $t=t_{0}$ shows well the graceful exit behaviour.
}
\label{figH}
\end{center}
\end{figure}

We can choose one Bohmian trajectory to see the evolution with
respect to time $t$. Figures \ref{figa} and \ref{figH} clearly show the
graceful exit behaviour, i.e., the evolution that begins with inflation
and smoothly changes to the decelerated expansion without any singularity in
the transition. The scale factor $a(t)$ and the Hubble parameter $H(t)$ of
the classical solutions are recovered when $\mid t-t_{0}\mid >>0$, but near
$t=t_{0}$ there is no singularity.

\begin{figure}
\begin{center}
\includegraphics[scale=0.75]{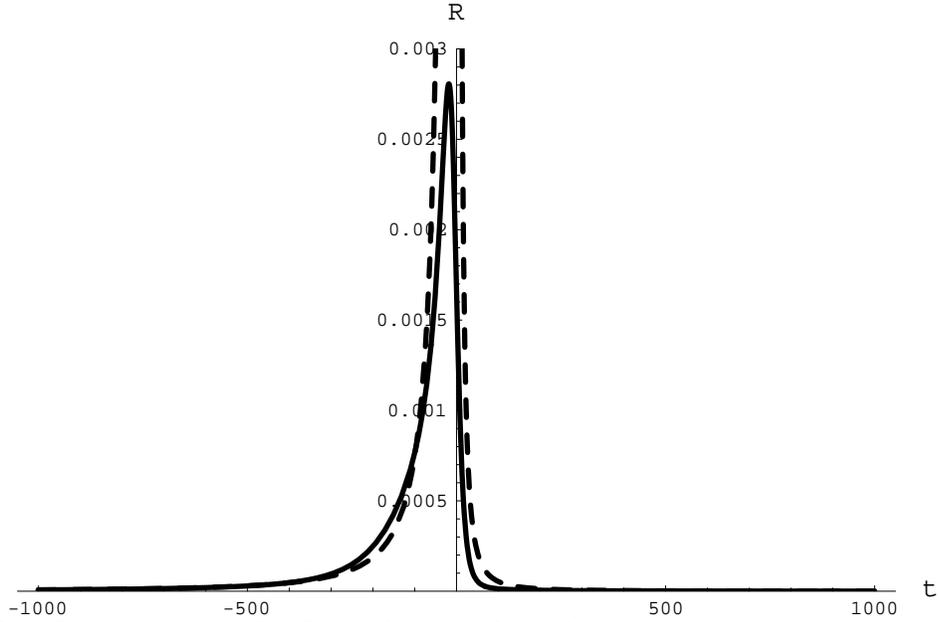}
\caption{
Plot of the Ricci curvature scalar $R(t)$ as function of $t$, with initial
conditions $\alpha =2$, $\phi =0$ and $t_{0}=0$ using the system of planar
equations (\ref{dalphadtJordan})--(\ref{dphidtJordan}) for $\sigma =d=1$. The
classical solutions of the $(+)$ and $(-)$ branches, shown in dashed lines,
are asymptotically obtained when $\mid t-t_{0}\mid >>0$, and the smooth
transition near the strong curvature regime, $t=t_{0}$, avoids the singularity
of the transition from inflationary to decelerated expansion.
}
\label{figR}
\end{center}
\end{figure}

For FLRW metrics with vanishing spatial curvature, the Ricci scalar
curvature $R$ reads
\begin{equation}
R=6\left[ \left( \frac{\dot{a}}{a}\right) ^{2}+\left( \frac{\ddot{a}}{a}%
\right) \right] =6\left( 2\dot{\alpha}^{2}+\ddot{\alpha}\right) =6\left(
2H^{2}+\dot{H}\right) ,  \label{R}
\end{equation}
where $H=\dot{a}/a=\dot{\alpha}$. Figure \ref{figR} shows the values of
$R$ for the Bohmian solution of Figure \ref{figa}. It can be seen that
it can be large in the quantum regime but it never diverges, exhibiting
a graceful exit transition.

\section{Conclusion}
\label{secConclusion}

We have shown that the Bohm-de Broglie ontological interpretation of
quantum cosmology applied to string cosmology can yield a simple answer
to the graceful exit problem in superinflation.

Using a previous study \cite{Fla3a1} in the Einstein frame, where
Gaussian superpositions of the solutions of the Wheeler-DeWitt equation
in the minisuperspace allows the determination of Bohmian trajectories,
we applied a conformal transformation to obtain in a straightforward way
the Bohmian trajectories in the Jordan frame corresponding to universe
histories in string cosmology.

We exhibited a class of trajectories representing universe evolutions
with graceful exit. They start classically from inflationary behaviour
driven by kinetic terms, as it is usual in these cases. When curvature
becomes relevant, quantum effects become important avoiding the
singularity and creating a smooth transition to the usual classical
decelerated expansion.

The results of this paper should be contrasted with results obtained
using other interpretations (see, e.g., Ref. \cite{kiefer}). In the case
of the Many Worlds interpretation, the construction of a Hilbert space
is very important in order to define a probability measure. However, for
scalar field minisuperspaces, the Wheeler-DeWitt equation has a
Klein-Gordon structure and it is well known that it is hard to construct
a unique positive definite measure with such equations. The common way
out from these problems is to take semiclassical approximations. This
problem is intimately connected with the problem of time \cite{kuchar}.
Once we do not have an extrinsic time, the Wheeler-DeWitt equation
cannot be transformed into a Schroedinger like equation, and hence all
these problems with its hyperbolic signature appear, unless one makes
use of a semiclassical approximation as it is done in Ref.
\cite{kiefer}. However, the graceful exit shall occur when quantum
effects are important, where a semiclassical approximation breaks down,
rendering problematic the analysis of this transition in the Many Worlds
interpretation. In the Bohm-de Broglie interpretation, probability
measures are not essential. We can talk about quantum trajectories, and
probabilities are relevant only for settling reasonable initial
conditions on such trajectories. The notion of quantum trajectories is
well defined even for Klein-Gordon like Wheeler-DeWitt equations. The
scale factor and scalar field follow real trajectories which satisfy
modified Einstein's equations ammended with a quantum potential term.
The concept of spacetime is as meaningfull in the Bohm-de Broglie
interpretation of minisuperspace quantum cosmology as it is in classical
General Relativity (GR) (see Ref. \cite{bola2} for a detailed discussion
on this point, and \cite{santini} for the different case of
midisuperspace models and the full superspace). The only difference is
that the dynamics may be changed by the presence of the quantum
potential. As in classical minisuperspace GR, the dynamics is invariant
under time reparametrizations, even in the presence of the quantum
potential. Hence, the symbol $t$ in the Bohmian trajectories is an
arbitrary parameter like in GR, which can be fixed by a choice of the
lapse function, without physical modifications of the dynamics of
spacetime. That is how it is possible, in this interpretation, to talk
about a graceful exit transition when the curvature of spacetime
increases. We have only to examine the behaviour of the Bohmian
trajectories in such regions. Note that in the region of configuration
space where a semiclassical description can be applied in the framework
of the Many Worlds interpretation, there is accordance of results: the
semiclassical wave function is peaked along the classical trajectories
\cite{kiefer}, which are exactly the Bohmian trajectories in that
region.

The conclusion is that, as the Bohm-de Broglie interpretation has more
concepts then the Many Worlds interpretation, the former can be used in
situations where the application of the latter is problematic. Our
results come in this way. Of course it would be desirable to define a
probability measure on the Bohmian trajectories coming from the wave
function (\ref{onda}), which could be valid in all minisuperspace, as
long as not all of them describe the Pre-Big Bang model with graceful
exit, although all Bohmian trajectories with initial conditions imposing
an initial superinflationary Pre-Big Bang phase present a graceful exit
to the decelerated regime. If this is possible, one should expect that
the statistical predictions of the Many Worlds interpretation are
identical to the statistical predictions of the Bohm-de Broglie
interpretation.

The above considerations raise doubts on the physical equivalence of
these two interpretations in quantum cosmology. Due to the singular
features present in the quantization of the whole Universe, it is
possible that one of them presents more physical relevant and testable
information then the other in this particular field (which is not the
case, up to now, in quantum field theory and non relativistic quantum
mechanics). This suggests that quantum cosmology may be the arena where
it could be possible to distinguish physically between these two
interpretations. This is a strong motivation to push forward these two
interpretations up to their limits in all possible aspects of quantum
cosmology and compare their results in order to obtain a relevant
physical distinction, or even some observable prediction, which could
select one of them as the valid interpretation of quantum mechanics.

A logical future generalization of the present work is to study dilaton
field potential terms so that the decelerated expansion can be joined
smoothly to an ordinary FRW radiation dominated expanding Universe
\cite{brustein}. It is also important to check how these results depend
on initial conditions on the WDW equation.

\acknowledgments

We would like to thank CNPq and CAPES of Brazil for financial support.
One of us (NPN) would like to thank the Laboratoire de Gravitation et
Cosmologie Relativistes of Universit\'e Pierre et Marie Curie where part
of this work has been done, for financial aid and hospitality, and the
group of ``Pequeno Semin\'ario'' in CBPF for useful discussions.

\newpage

\end{document}